\documentclass[aps,pra,twocolumn,reprint]{revtex4-2}
\usepackage{graphicx}
\usepackage[english]{babel}
\usepackage[inline]{enumitem}
\usepackage{amsmath, amssymb, amsfonts}
\usepackage[caption=false,labelformat=simple]{subfig}% For subfigure
\usepackage{bm}
\usepackage{xcolor}
\usepackage[colorlinks=true,linkcolor=blue]{hyperref}
\usepackage[normalem]{ulem}

\newlength{\figwidth}
\newlength{\lfig}
\newlength{\sfig}
\begin{document}
\title{Controlling collisional loss and scattering lengths\break
of ultracold dipolar molecules with static electric fields}

\author{Bijit Mukherjee}
\affiliation{Joint Quantum Centre (JQC) Durham-Newcastle, Department of Chemistry, Durham University, South Road, Durham, DH1 3LE, United Kingdom.}
\author{Jeremy M. Hutson}
\email{j.m.hutson@durham.ac.uk} \affiliation{Joint Quantum Centre (JQC)
Durham-Newcastle, Department of Chemistry, Durham University, South Road,
Durham, DH1 3LE, United Kingdom.}

\date{\today}

\begin{abstract}
Trapped samples of ultracold molecules are often short-lived, because close collisions between them result in trap loss. We investigate the use of shielding with static electric fields to create repulsive barriers between polar molecules to prevent such loss. Shielding is very effective even for RbCs, with a relatively low dipole moment, and even more effective for molecules such as NaK, NaRb and NaCs, with progressively larger dipoles. Varying the electric field allows substantial control over the scattering length, which will be crucial for the stability or collapse of molecular Bose-Einstein condensates. This arises because the dipole-dipole interaction creates a long-range attraction that is tunable with electric field. For RbCs, the scattering length is positive across the range where shielding is effective, because the repulsion responsible for shielding dominates. For NaK, the scattering length can be tuned across zero to negative values. For NaRb and NaCs, the attraction is strong enough to support tetraatomic bound states, and the scattering length passes through resonant poles where these states cross threshold. For KAg and CsAg, there are multiple bound states and multiple poles. For each molecule, we calculate the variation of scattering length with field and comment on the possibilities for exploring new physics.
\end{abstract}

\maketitle

\section{Introduction}

Trapped samples of ultracold polar molecules offer opportunities to study important physical phenomena that range from quantum simulation \cite{Baranov:2012, Blackmore:2019} to quantum magnetism \cite{Gorshkov:2011, Gadway:2016}. However, close collisions between dipolar molecules often result in trap loss due to short-range processes such as chemical reaction \cite{Ospelkaus:react:2010}, inelastic collisions, 3-body recombination \cite{Mayle:2013} or laser excitation of collision complexes \cite{Christianen:laser:2019}. Because of this, there is much interest in shielding molecules from close collisions by engineering long-range barriers between them. This may be done with static electric fields \cite{Avdeenkov:2006, Wang:dipolar:2015, Quemener:2016, Gonzalez-Martinez:adim:2017, Mukherjee:CaF:2023}, near-resonant microwaves \cite{Karman:shielding:2018, Karman:shielding-imp:2019, Lassabliere:2018}, or lasers \cite{Xie:optical:2020}. Both microwave shielding and static-field shielding have been demonstrated experimentally \cite{Matsuda:2020, Li:KRb-shield-3D:2021, Anderegg:2021}, and microwave shielding has recently been used to achieve Fermi degeneracy for fermionic NaK \cite{Schindewolf:NaK-degen:2022}. Bose-Einstein condensation has not yet been achieved for dipolar molecules, but is the subject of active experimental efforts \cite{Bigagli:NaCs:2023, Lin:NaRb:2023}.

Shielding relies on the dipole-dipole interaction between two polar molecules. This is proportional to $1/R^3$ at long range, where $R$ is the intermolecular distance. Two different states of a molecular pair are tuned into near-degeneracy with an electric, microwave or laser field. If the two pair states are connected by the dipole-dipole interaction, they repel one another. It is thus possible to engineer an effective interaction potential for the upper pair state that has a repulsive component at moderately long range.

We have recently developed an efficient computational approach to simulating shielding with static fields \cite{Mukherjee:CaF:2023}. This is based on coupled-channel calculations of inelastic processes, with an absorbing boundary condition \cite{Clary:1987, Janssen:PhD:2012} to take account of short-range loss without simulating it in detail. Our approach is conceptually similar to earlier methods for static-field \cite{Wang:dipolar:2015} and microwave \cite{Karman:shielding:2018} shielding, but is made far more efficient by using a basis set of field-dressed molecular functions and by taking account of energetically well-separated functions using a Van Vleck transformation \cite{VanVleck:1928, Kemble:1937}. We have applied this approach to shielding of CaF, which we found to be very effective at electric fields of approximately 23 kV/cm.

Ultracold collisions are often characterized by a scattering length $a$ \cite{Hutson:theory-cold-colls:2009, Chin:RMP:2010}, which describes the phase shift for completed low-energy collisions. The scattering length governs both 2-body and many-body properties, and in many-body physics is often represented by a zero-range contact interaction \cite{Huang:1957}. For dipolar systems, many-body properties arise from an interplay between the scattering length and long-range anisotropic interactions; the latter depend on the space-fixed dipole moment $d$ and may be affected by the geometry of the system. The mean-field ansatz usually adopted \cite{Yi:2000, Yi:2001} involves supplementing a contact potential based on an effective scattering length $a_\textrm{eff}$ with a long-range dipole-dipole interaction based on $d$. This approach has been validated by Ronen \emph{et al.}\ \cite{Ronen:2006}, who compared the results of the mean-field prescription with those of fully correlated calculations. An important point is that $a_\textrm{eff}$ is affected by long-range forces and depends on $d$, so Ronen \emph{et al.}\ referred to it as the \emph{dipole-dependent} scattering length.

In the absence of a long-range well, the repulsion due to shielding would be expected to produce a positive scattering length. However, we have previously shown \cite{Mukherjee:CaF:2023} that, for CaF, the scattering length is large and negative in the region where shielding is effective. This arises because there is in fact a long-range well outside the shielding barrier; this adds a negative contribution that overcomes the positive one from shielding.

In this paper, we explore the tunability of scattering lengths for a variety of polar molecules shielded with static electric fields and outline the scientific possibilities that this offers. We show that the scattering length depends on the applied field in a non-trivial way, and not just on the dipole moment $d$ induced by the field. Indeed, the same scattering length can often be achieved for different fields that induce different dipole moments. This opens the way to investigating many-body properties as independent functions of scattering length and dipole moment.

A wide range of behavior is possible for different molecules. For some molecules, such as RbCs, the long-range attractive well is shallow and the scattering length has moderate positive values. For others, such as NaK, the attractive well is deeper and the scattering length can be tuned to zero; this may allow the production of a quantum gas with purely dipolar interaction. There are also some systems, such as NaRb and NaCs, where the attractive well is deep enough to support one or more bound states; in this case, adjusting the electric field across the threshold for creation of a bound state allows ``dialling up" any desired scattering length. It may also allow ``electroassociation" to form long-range tetraatomic molecules.

The structure of the paper is as follows. Section \ref{sec:theory} describes our theoretical approach, including the Hamiltonians and basis-set expansions used, and describes how the properties of individual molecules affect shielding and the ability to control scattering lengths. Section \ref{sec:results} presents our results for each of the representative molecules mentioned above. Finally, Section \ref{sec:conclusions} presents conclusions and perspectives.

\section{Theory}
\label{sec:theory}

\subsection{Coupled-channel approach}

\begin{figure}[tbp]
\begin{center}
\includegraphics[width=\figwidth,clip]{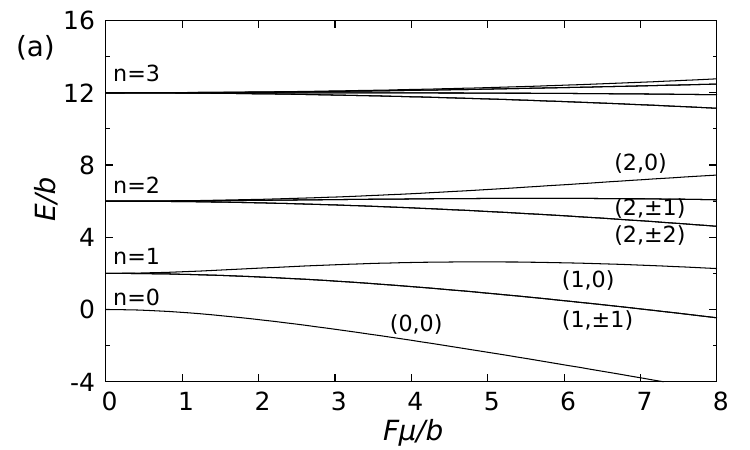}
\includegraphics[width=\figwidth,clip]{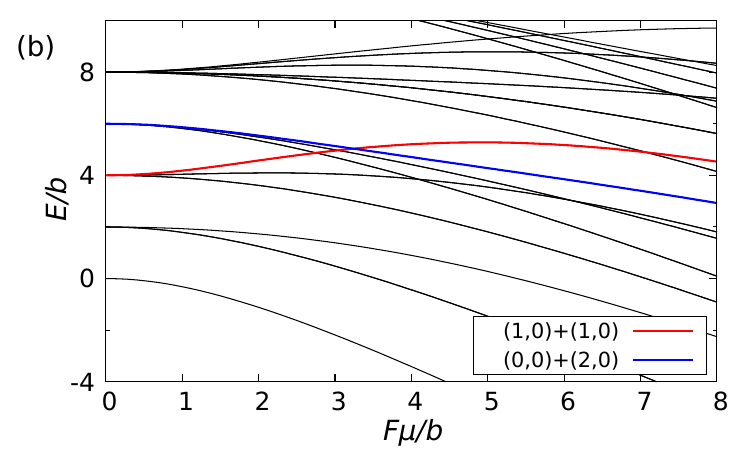}
\caption{Energy of (a) a single molecule in a $\Sigma$ state and (b) a pair of molecules as a function of electric field. The initial pair state that can be shielded from close collisions, ($\tilde{n},m_n)$ = (1,0)+(1,0), is shown in red. The state (0,0)+(2,0) that is responsible for shielding is shown in blue, and crosses the initial state near $F\mu/b = 3.244$.}
\label{fig:mol_Stark}
\end{center}
\end{figure}

The theoretical approach used here is similar to that in ref.\ \cite{Mukherjee:CaF:2023}.
If vibrational motion and spins are neglected, the Hamiltonian for a single linear polar molecule $k$ in a $\Sigma$ state, in an electric field $\boldsymbol{F}$, is
\begin{equation}
\hat{h}_k = b_k\hat{\boldsymbol{n}}_k^2 - \boldsymbol{\mu}_k \cdot \boldsymbol{F},
\label{eq:ham-Stark}
\end{equation}
where $\hat{\boldsymbol{n}}_k$ is the operator for molecular rotation, $b_k$ is the rotational constant, and $\boldsymbol{\mu}_k$ is the dipole moment along the molecular axis. It is convenient to represent the resulting energy levels in reduced units, $E/b$ for energy and $F\mu/b$ for field, so that they can be applied to any molecule in a $\Sigma$ state. Figure~\ref{fig:mol_Stark}(a) shows the single-molecule energy levels as a function of electric field; we label the levels $(\tilde{n},m_n)$; here $\tilde{n}$ is a quantum number that correlates with the free-rotor quantum number $n$ at zero field and $m_n$ represents the conserved projection of $n$ onto the space-fixed $z$ axis. Figure~\ref{fig:mol_Stark}(b) shows the corresponding energy of a \emph{pair} of noninteracting molecules. Shielding can occur when the dipole-dipole interaction connects two pair states that are close together in energy. In Figure~\ref{fig:mol_Stark}(b), it can occur when two molecules in the state (1,0) collide at fields just above $F\mu/b = 3.244$, where (1,0)+(1,0) lies just above (0,0)+(2,0). Shielding is most effective at reduced field $F\mu/b \approx 3.4$, when the space-fixed dipole moment $d$ of the state (1,0) is about $-0.13\mu$ \cite{Gonzalez-Martinez:adim:2017}.

The Hamiltonian for a colliding pair of molecules is
\begin{equation}
\hat{H} = \frac{\hbar^2}{2\mu_\textrm{red}}\left( -R^{-1} \frac{d^2}{dR^2} R \right)
+ \hat{H}_\textrm{int},
\label{eq:ham-pair}
\end{equation}
where $R$ is the intermolecular distance and $\mu_\textrm{red}$ is the reduced mass. The internal Hamiltonian for the pair is
\begin{equation}
\hat{H}_\textrm{int} = \hat{h}_1 + \hat{h}_2 + \hat{H}_\textrm{cent} + V_\textrm{int},
\end{equation}
where
\begin{equation}
\hat{H}_\textrm{cent}=\frac{\hbar^2\hat{\boldsymbol{L}}^2}{2\mu_\textrm{red}R^2}
\end{equation}
is the centrifugal Hamiltonian and $\hat{\boldsymbol{L}}^2$ is the angular momentum operator for rotation of the two molecules about one another. The interactions involved in shielding occur at long range, so here we approximate the interaction potential $V_\textrm{int}$ by the dipole-dipole operator,
\begin{equation}
\hat{H}_\textrm{dd} = -[3(\boldsymbol{\mu}_1\cdot\hat{\boldsymbol{R}}) (\boldsymbol{\mu}_2\cdot\hat{\boldsymbol{R}}) - \boldsymbol{\mu}_1\cdot\boldsymbol{\mu}_2] / (4\pi\epsilon_0 R^3),
\label{eq:Hdd}
\end{equation}
where $\hat{\boldsymbol{R}}$ is a unit vector along the intermolecular axis.

The total wavefunction is expanded as
\begin{equation} \Psi(R,\hat{\boldsymbol{R}},\hat{\boldsymbol{r}}_1,\hat{\boldsymbol{r}}_2)
=R^{-1}\sum_j\Phi_j(\hat{\boldsymbol{R}},\hat{\boldsymbol{r}}_1,\hat{\boldsymbol{r}}_2)\psi_{j}(R), \label{eq:expand}
\end{equation}
where $\hat{\boldsymbol{r}}_k$ is a unit vector along the axis of molecule $k$.
We use a basis set of functions $\{\Phi_j\}$,
\begin{equation}
\Phi_j = \phi^{\tilde{n}_1}_{m_{n1}}(\hat{\boldsymbol{r}}_1) \phi^{\tilde{n}_2}_{m_{n2}}(\hat{\boldsymbol{r}}_2) Y_{LM_L}(\hat{\boldsymbol{R}}),
\label{eq:basis-spin-free}
\end{equation}
symmetrized for exchange of identical bosons. Here $\phi^{\tilde{n}_1}_{m_{n1}}(\hat{\boldsymbol{r}}_1)$ and $\phi^{\tilde{n}_2}_{m_{n2}}(\hat{\boldsymbol{r}}_2)$ are field-dressed rotor functions that diagonalize $\hat{h}_1$ and $\hat{h}_2$, respectively, and $Y_{LM_L}(\hat{\boldsymbol{R}})$ are spherical harmonics that are the eigenfunctions of $\hat{\boldsymbol{L}}^2$. The field-dressed functions $\phi^{\tilde{n}}_{m_{n}}(\hat{\boldsymbol{r}})$ are themselves expanded in free-rotor functions $Y_{nm_n}(\hat{\boldsymbol{r}})$.

Substituting the expansion (\ref{eq:expand}) into the total Schr\"odinger equation produces a set of coupled equations in the intermolecular distance $R$. These equations are solved separately for each value of the projection of the total angular momentum, $M_\textrm{tot}=m_{n1}+m_{n2}+M_L$, which is a conserved quantity.
The eigenvalues of $\hat{H}_\textrm{int}$ as a function of $R$ form a set of adiabats $U_i(R)$ that represent effective potentials for relative motion. The computational method of ref.\ \cite{Mukherjee:CaF:2023} does not rely on these adiabats and takes full account of transitions between them, but the adiabats are nevertheless conceptually valuable tools for understanding the scattering properties.

The dipole-dipole operator is off-diagonal in the partial-wave quantum number $L$, and converged calculations typically require substantial values of $L_\textrm{max}$. The basis sets can thus become very large if substantial numbers of pair functions $(\tilde{n}_1,m_{n1},\tilde{n}_2,m_{n2})$ are included explicitly. The problem becomes extreme when electron and nuclear spins are included. We therefore include only a relatively small number of ``class 1" pair functions (with labels a, b, $\ldots$) explicitly in the basis set, with the remaining ``class 2" functions (with labels $\alpha$, $\beta$, $\ldots$) taken into account through a Van Vleck transformation as described in ref.\ \cite{Mukherjee:CaF:2023}. The transformation contributes matrix elements \emph{between} the channels in class 1 of the form
\begin{align}
\langle a &| \hat{H}_\textrm{dd,VV} | b \rangle \nonumber\\
&= \sum_\alpha \frac{1}{2} \left[
\frac{\langle a | \hat{H}_\textrm{dd} | \alpha \rangle
\langle \alpha | \hat{H}_\textrm{dd} | b \rangle} {(E_a-E_\alpha)}
+ \frac{\langle a | \hat{H}_\textrm{dd} | \alpha \rangle
\langle \alpha | \hat{H}_\textrm{dd} | b \rangle} {(E_b-E_\alpha)}
\right].
\end{align}
We make the further approximation of replacing the energies in the denominators with their asymptotic values, so that they are independent of $R$ and $\hat{H}_\textrm{dd,VV}$ is proportional to $R^{-6}$.

We use a basis set with $\tilde{n}$ up to 5 to calculate the monomer properties. For the pair functions, the Van Vleck treatment allows us to restrict the basis set of functions in class 1 to those with $\tilde{n}\le 2$, and to include the remaining functions up to $\tilde{n}=5$ in class 2. In this way, we achieve an enormous reduction in computational cost without compromising the convergence~\cite{Mukherjee:CaF:2023}. We include functions with even $L$ up to $L_\textrm{max}=20$.

\subsection{Cross sections and scattering lengths}

Colliding molecules may be lost from a trap in two ways. First, colliding pairs may undergo a transition to a lower-lying pair state. This releases kinetic energy that is usually sufficient to eject both molecules from the trap, and we refer to it as inelastic loss. Secondly, any pairs that penetrate through the engineered repulsive barrier and reach small intermolecular distances are also likely to be lost, through short-range inelasticity, laser absorption, or 3-body collisions. We refer to this as short-range loss and to the sum of inelastic and short-range loss as total loss.

To model these processes, we solve the coupled equations with a fully absorbing boundary condition at short range \cite{Clary:1987, Janssen:PhD:2012}. The numerical methods used are as described in ref.\ \cite{Mukherjee:CaF:2023}. This is done separately for each electric field $F$ and collision energy $E_\textrm{coll}$, producing a non-unitary S matrix $\boldsymbol{S}$ and cross sections $\sigma_\textrm{el}$, $\sigma_\textrm{inel}$ and $\sigma_\textrm{short}$ for elastic scattering, inelastic scattering and short-range loss, respectively. The corresponding rate coefficients $k$ at collision energy $E_\textrm{coll}$ are related to the cross sections $\sigma$ through $k = v \sigma$, where $v=(2E_\textrm{coll}/\mu_\textrm{red})^{1/2}$. We also obtain a complex energy-dependent scattering length \cite{Hutson:res:2007},
\begin{equation}
a(k_0) = \alpha(k_0)-i\beta(k_0) = \frac{1}{ik_0} \left(\frac{1-S_{00}(k_0)}{1+S_{00}(k_0)}\right),
\label{eq:a-k}
\end{equation}
where $k_0=(2\mu_\textrm{red} E_\textrm{coll}/\hbar^2)^{1/2}$.

The interaction potential between two polar molecules is typically deep and strongly anisotropic at short range. However, shielding occurs due to dipole-dipole interactions that occur at intermolecular distances $R\gg 100\ a_0$. At these distances, the chemical interactions that dominate at short range make very little contribution, and they are neglected in the present work. However, there are also effects due to dispersion interactions, which are proportional to $R^{-6}$ at long range. These are of two types. Rotational dispersion interactions arise from matrix elements of $\hat{H}_\textrm{dd}$ off-diagonal in monomer rotational quantum numbers, which are included directly in the coupled equations. In addition, there are electronic dispersion interactions, arising from dipole-dipole matrix elements off-diagonal in electronic state. We take these into account through an additional interaction $V_\textrm{disp}^\textrm{elec} = -C_6^\textrm{elec}/R^6$. This makes only a small contribution to the loss rate, at the level of 10\% for RbCs but much less for the other molecules considered here.

\subsection{Reduced units}

Dipole-dipole scattering may be cast in a dimensionless form, with lengths expressed in terms of a dipole length $R_3 = (2\mu_\textrm{red}/\hbar^2) (\mu^2/4\pi\epsilon_0)$ and energies expressed in terms of a corresponding energy $E_3 = \hbar^2/(2\mu_\textrm{red} R_3^2)$ \cite{Gonzalez-Martinez:adim:2017}. A crucial quantity for shielding with a static field is the reduced rotational constant of the molecule, $\tilde{b}=b/E_3$. For ultracold molecules of current interest, this ranges from ${\sim}10^6$ for NaLi to ${\sim}10^{11}$ for YO \cite{Gonzalez-Martinez:adim:2017}, and is even larger for some molecules containing Ag. Static-field shielding is generally most effective for molecules with larger values of $\tilde{b}$. Table \ref{tab:reduced} summarizes the key quantities for the molecules considered in the present work.

\begin{table*}[tbp]
\caption{Key quantities for the molecules considered in this study. Molecular dipole moments and rotational constants are taken from refs.\  \cite{Molony:RbCs:2014, Gregory:RbCs-microwave:2016, Gerdes:2011,Russier-Antoine:2000, Guo:NaRb:2016, Guo:NaRb:2018, Dagdigian:1972, Docenko:2006, Smialkowski:2021}.
\label{tab:reduced}} \centering
\begin{ruledtabular}
\begin{tabular}{cccccc}
%\hline\hline
& $b/\mu$ (kV/cm) & $R_3$ ($a_0$)  & $E_3/k_\textrm{B}$ (K) & $\tilde{b}$ &
$C_6^\textrm{elec}/(E_3R_3^6)$\footnotemark\footnotetext[1]{Values of $C_6^\textrm{elec}$ are taken from Ref. \cite{Lepers:2013}.} \\ \hline
$^{87}$Rb$^{133}$Cs  &  0.795 &  $9.3 \times 10^4$ &  $9.1 \times 10^{-11}$  &  $2.6 \times 10^8$  &  $9.5 \times 10^{-11}$ \\
$^{23}$Na$^{39}$K   &  2.085 & $1.3 \times 10^5$ & $1.7 \times 10^{-10}$ & $8.2 \times 10^8$  &  $3.0 \times 10^{-12}$ \\
$^{23}$Na$^{87}$Rb   &  1.297  & $3.2 \times 10^5$ & $1.6 \times 10^{-11}$ & $6.4 \times 10^9$  &  $1.8 \times 10^{-13}$ \\
$^{23}$Na$^{133}$Cs   &  0.726  & $9.9 \times 10^5$ & $1.1 \times 10^{-12}$ & $7.4 \times 10^{10}$  &  $3.5 \times 10^{-15}$ \\
KAg
\footnotemark\footnotetext[2]{Weighted-average atomic masses are used.}
 &  0.467  & $3.0 \times 10^6$ & $1.3 \times 10^{-13}$ & $7.3 \times 10^{11}$  &  $3.3 \times 10^{-17}$
\footnotemark\footnotetext[3]{Calculated $C_6^\textrm{elec}$ is not available, so we use a value of 10000 $E_\textrm{h}a_0^6$.} \\
CsAg \footnotemark[2]  &  0.164 & $6.5 \times 10^6$ & $1.7 \times 10^{-14}$ & $2.2 \times 10^{12}$  &  $2.5 \times 10^{-18}$~\footnotemark[3] \\
%\hline\hline
\end{tabular}
\end{ruledtabular}
\end{table*}

\subsection{Adiabatic curves}
\label{sec:adiabatic}

\begin{figure*}[tbp]
	\subfloat[]{
		\includegraphics[width=0.45\textwidth]{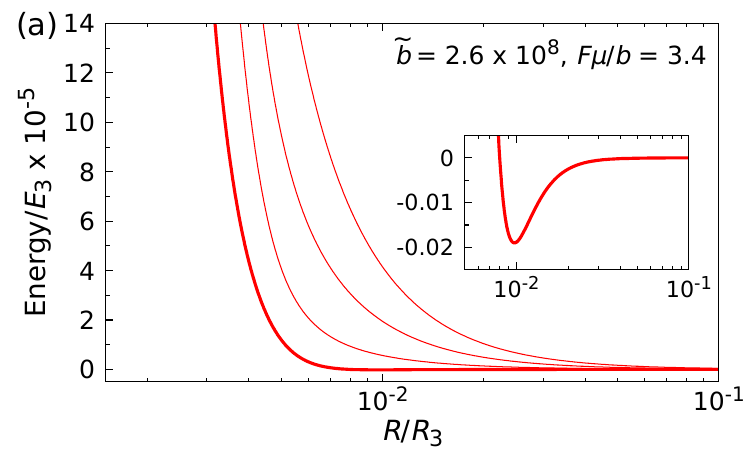}
	}
	\subfloat[]{
		\includegraphics[width=0.45\textwidth]{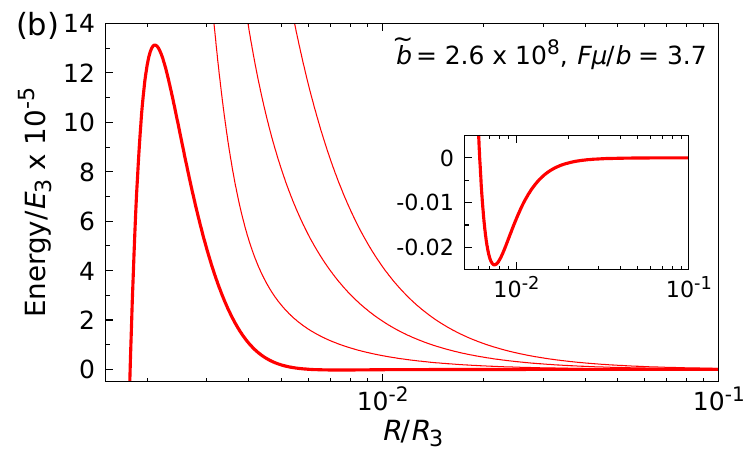}
	}
    %%%%%%%%%%%%%%%%%%%%%%%%%%%%%%%%%%%%second row
    \vspace{-1 cm}
	\subfloat[]{
		\includegraphics[width=0.45\textwidth]{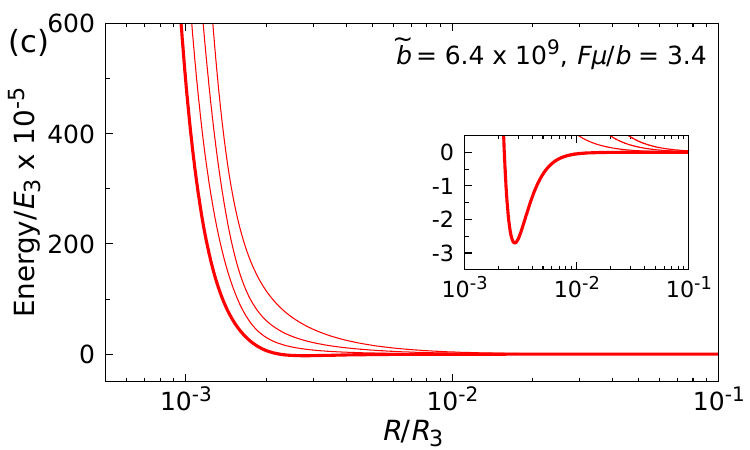}
	}
	\subfloat[]{
		\includegraphics[width=0.45\textwidth]{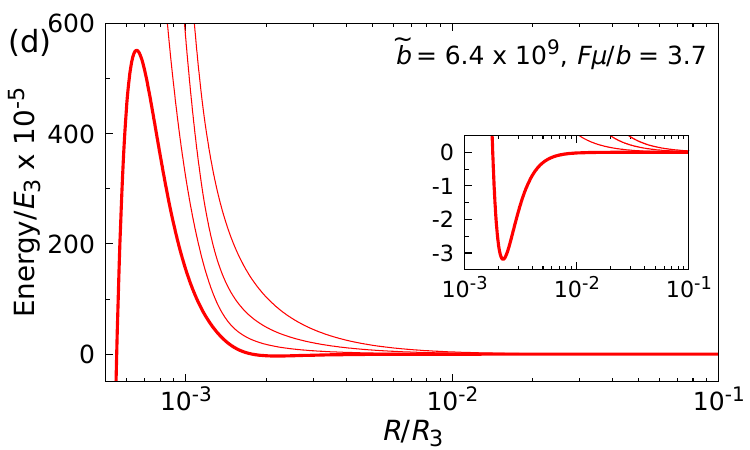}
	}
    \vspace{-0.5 cm}
    \caption{Adiabats correlating asymptotically to partial waves $L=0,2,4,6$ at the incoming threshold (1,0)+(1,0). The thick lines in each panel is the adiabat for the s-wave channel. Panels (a) and (b) show adiabats for a value of $\tilde{b}$ appropriate for RbCs at two different values of the reduced electric field $F\mu/b$. Panels (c) and (d) show adiabats for a larger $\tilde{b}$ (appropriate for NaRb) with the same two values of $F\mu/b$. The insets show expanded views of the attractive region of the incoming s-wave channel.
}%
\label{fig:adiabats}
\end{figure*}

Figure \ref{fig:adiabats} shows the adiabats correlating with the incoming pair state (1,0)+(1,0), in reduced units, for values of $\tilde{b}$ appropriate for NaRb and RbCs. The adiabats are shown for reduced electric fields of 3.4 (where shielding is typically most effective) and 3.7 (where shielding is less effective for RbCs). When (0,0)+(2,0) lies slightly below (1,0)+(1,0), as here, the adiabats for (1,0)+(1,0) are repulsive at distances of a few hundred bohr due to mixing with the lower threshold. However, the adiabat for s-wave scattering ($L=0$) is attractive at larger distances. This arises because $\hat{H}_\textrm{dd}$ mixes channels with $\Delta L=\pm2$ arising from the same threshold; since $\hat{H}_\textrm{dd}$ is proportional to $R^{-3}$ and the separation between such channels is proportional to $R^{-2}$, this produces an attraction of the form $-C_4/R^4$ at long range, with coefficient $C_4 = (2/15) (d/\mu)^4 E_3R_3^4$, where $d$ is the space-fixed dipole induced by the electric field.

In reduced units, the coefficient of the long-range attraction is the same for any polar molecule at the same reduced field. However, the repulsion depends on the separation of rotational levels, and in reduced units the leading term is proportional to $\tilde{b}^{-1}$. Because of this, the attraction persists to shorter reduced distances for larger $\tilde{b}$, and the resulting well is deeper in reduced energy units. Figure \ref{fig:adiabats} shows that, in reduced units, it is 2 orders of magnitude deeper for NaRb than for RbCs. This is the origin of the molecule-dependent effects on scattering length described below. However, it should be noted that the effects on range and depth are often reversed in absolute units.

The height of the repulsive barrier also depends on $\tilde{b}$ and reduced field. The finite height arises because there are additional adiabats that come down from higher thresholds and undergo avoided crossings with those from (1,0)+(1,0). The top of the barrier is the lower limb of such an avoided crossing. The relative energies of the higher thresholds, in reduced units, are proportional to $\tilde{b}$ and decrease as a function of reduced field. The barrier maximum is thus at larger reduced distance for smaller $\tilde{b}$ or larger reduced field, and the barrier is correspondingly lower. Figure \ref{fig:adiabats} shows that, in reduced units, the barrier is about a factor of 40 higher for NaRb than for RbCs at $F\mu/b = 3.7$. This is responsible for the more effective shielding for larger $\tilde{b}$ described below.

\section{Results}
\label{sec:results}

This section describes the results of coupled-channel calculations of shielded collisions for a variety of systems. We consider a sequence of bosonic alkali-metal diatomic molecules, in order of increasing $\tilde{b}$. This corresponds to increasing depth of the long-range well when expressed in reduced energy units. We consider both the effectiveness of shielding and the elastic scattering properties, including the real part $\alpha$ of the scattering length $a$. For many-body properties, $\alpha$ will play the same role as the dipole-dependent scattering length of Ronen \emph{et al.}\ \cite{Ronen:2006}. In each case we identify interesting new physics that is made accessible by the special properties of the system concerned.

We use a fixed collision energy $E_\textrm{coll}=10$ nK$\times k_\textrm{B}$ in all calculations in the present work. This is a reasonable lower limit for the temperatures that are likely to be experimentally accessible, and where Bose-Einstein condensation may occur.
It is nevertheless quite large compared to the characteristic energies of dipolar collisions of molecules with high $\tilde{b}$, such as NaCs and the Ag systems; at the fields where shielding is most efficient, the space-fixed dipole $d$ is about $-0.13\mu$, so the characteristic energy is about $(\mu/d)^4 E_3\approx 3500E_3$ \cite{Gonzalez-Martinez:adim:2017}.

The rate coefficients for inelastic and short-range loss are generally dominated by s-wave scattering; they vary weakly or decrease with collision energy between 10~nK and 1~$\mu$K. Those for elastic scattering have an s-wave contribution that may depend strongly on the field, together with substantial contributions from higher partial waves; both contributions are proportional to $E_\textrm{coll}^{1/2}$ at low energy, though the s-wave contribution may drop below this at $E_\textrm{coll}>3500E_3$ \cite{Mukherjee:CaF:2023}. The ratio $\gamma$ of elastic scattering to loss thus increases with collision energy in most cases.

\subsection{RbCs}
\label{sec:RbCs}

\begin{figure}[tbp]
		\includegraphics[width=0.45\textwidth]{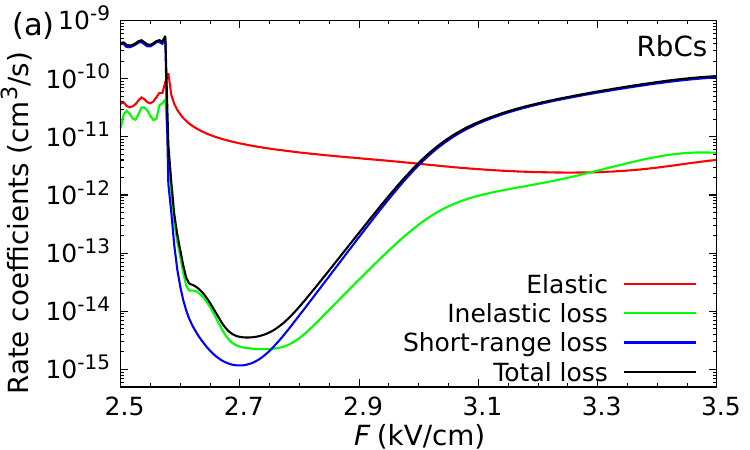}
		\includegraphics[width=0.45\textwidth]{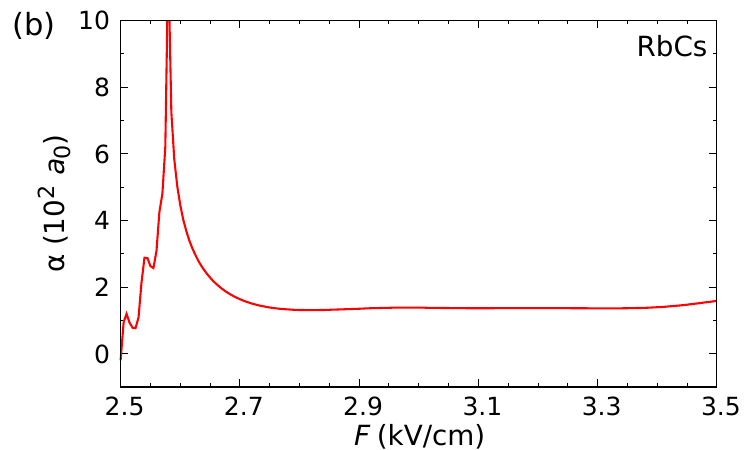}
    \caption{(a) Rate coefficients for two-body elastic collisions and loss processes for RbCs as a function of electric field, and (b) real part $\alpha(k_0)$ of scattering length calculated at collision energy $E_\textrm{coll}/k_\textrm{B}=10$\ nK.}%
    \label{fig:RbCs:results}
\end{figure}

Figure\ \ref{fig:RbCs:results}a shows the calculated rate coefficients for inelastic, short-range and total loss for RbCs as a function of electric field, in the shielding region, together with the rate coefficient for elastic collisions. The results shown are for $^{87}$Rb$^{133}$Cs, but other isotopologues would give results almost identical on the scale of the Figure. At a field of 2.7 kV/cm, the rate coefficient for total loss is more than three orders of magnitude lower than that for elastic collisions. At a density of $10^{13}$ cm$^{-3}$, plausible for a Bose-Einstein condensate (BEC), the calculated loss rate corresponds to a lifetime of around 30~s.

For shielded collisions, even elastic collisions are governed by shielding barriers that are mostly due to dipole-dipole forces and occur at long range. The scattering lengths are therefore predicted reliably by the current calculations. Figure \ref{fig:RbCs:results}b shows the real part $\alpha$ of the scattering length as a function of field. This is large and positive near 2.58 kV/cm, where (1,0)+(1,0) is nearly degenerate with (0,0)+(2,0); the resulting strong repulsion due to shielding produces a large positive contribution to $\alpha$, which decreases quickly above this field as the two pair states separate.

The attractive well for RbCs is not deep enough to support a bound state. Despite this, the scattering length is positive in the shielding region, with a value near 200 $a_0$ near 2.7 kV/cm. This arises because, in this system, the inner turning point of the adiabat for $L=0$ is near 750 $a_0$; the resulting excluded volume makes a positive contribution to the scattering length, which is larger than the negative contribution due to the long-range attraction. The overall value is favorable for sympathetic cooling, without being so large that the sample will enter the hydrodynamic regime at low densities. It is also favorable for producing a condensate that is stable (and does not collapse) over a wide range of geometries. Furthermore, the lack of a bound state outside the shielding barrier implies that 3-body recombination cannot occur. However, 3-body inelasticity (to produce separated molecules with one of them being in a lower rotational state) is still possible.

\subsection{NaK}
\label{sec:NaK}

\begin{figure}[tbp]
		\includegraphics[width=0.45\textwidth]{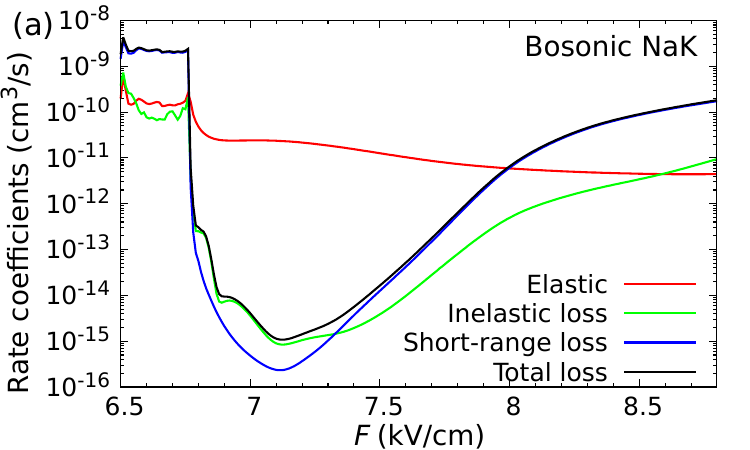}
		\includegraphics[width=0.45\textwidth]{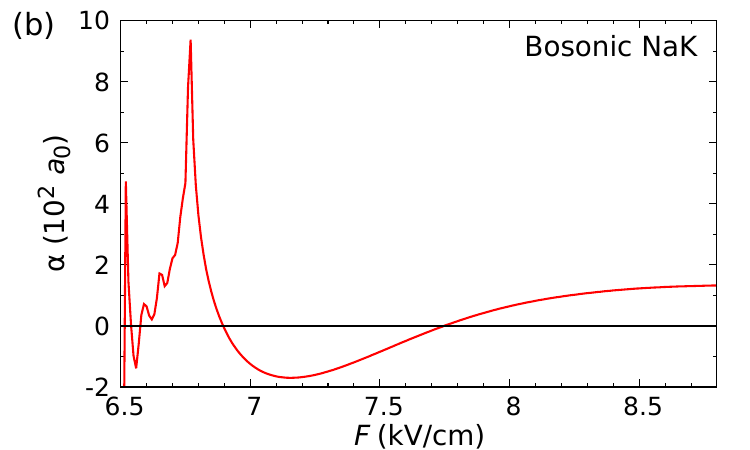}
    \caption{(a) Rate coefficients for two-body elastic collisions and loss processes for NaK as a function of electric field, and (b) real part $\alpha(k_0)$ of the scattering length calculated at collision energy $E_\textrm{coll}/k_\textrm{B}=10$\ nK.}%
    \label{fig:NaK:results}
\end{figure}

Figure\ \ref{fig:NaK:results}a shows the calculated rate coefficients for elastic scattering and loss for bosonic NaK. The results shown are for $^{23}$Na$^{39}$K, but $^{23}$Na$^{41}$K would give results almost identical on the scale of the Figure. At a field of 7.1 kV/cm, the rate coefficient for total loss is about four orders of magnitude lower than that for elastic collisions.

The real part $\alpha$ of the low-energy scattering length for NaK is shown in Figure \ref{fig:NaK:results}b. In this case $\alpha$ is negative at the center of the shielding region, indicating that the negative contribution from the long-range attraction is sufficient to overcome the positive contribution from the shielding core. However, this is not simply because the attraction is stronger for NaK than for RbCs. Indeed, in reduced units the attractive coefficient $C_4/(E_3 R_3^4)$ is the same for all systems at the same reduced field $F\mu/b$. However, as described in section \ref{sec:adiabatic}, the shielding repulsion is proportional to $\tilde{b}^{-1}$, so that the attraction dominates to lower reduced distance for larger $\tilde{b}$. For NaK at 7.1 kV/cm, the inner turning point of the lowest adiabat is at $R_\textrm{in}=0.005 R_3=650\ a_0$, compared to $0.008 R_3$ for RbCs at the same value of $F\mu/b$.

The negative value of $\alpha$ implies that condensates of NaK shielded by a static field of 7.1 kV/cm will probably be unstable with respect to collapse, except perhaps in strongly anisotropic traps dominated by side-to-side dipolar repulsions \cite{Ronen:2006}. However, the scattering length is controllable, with zeroes near 6.90 kV/cm and 7.75 kV/cm. Shielding is still effective at 6.90 kV/cm, with a ratio $\gamma\sim2500$ between the rate coefficients for elastic scattering and loss, but less effective at 7.75 kV/cm, with $\gamma\sim 25$. It may be noted that $\gamma$ usually increases approximately as $E_\textrm{coll}^{1/2}$ at low energy \cite{Quemener:2016}. The space-fixed dipole of NaK is $-0.38$ D at 6.90 kV/cm and $-0.29$ D at 7.75 kV/cm. This should permit the formation of dipolar condensates with positive or slightly negative scattering lengths, and should allow exploration of their stability as a function of geometry. The regions close to zeroes are particularly interesting, because they correspond to purely dipolar condensates with only weak non-dipolar interactions, so that the dipolar interactions will be particularly important.

\begin{figure}[tbp]
		\includegraphics[width=0.45\textwidth]{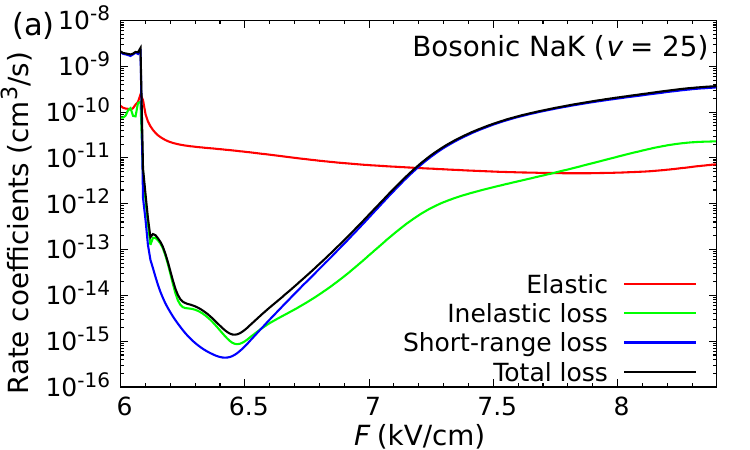}
		\includegraphics[width=0.45\textwidth]{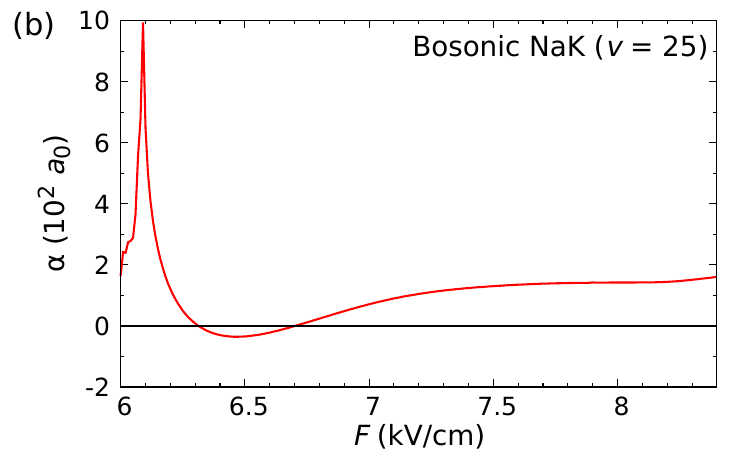}
    \caption{(a) Rate coefficients for two-body elastic collisions and loss processes for NaK ($v=25$) as a function of electric field, and (b) real part $\alpha(k_0)$ of scattering length calculated at collision energy $E_\textrm{coll}/k_\textrm{B}=10$\ nK.}%
    \label{fig:NaK-scaled:results}
\end{figure}

A further possibility arises from tuning the molecular dipole moment itself. Vibrationally excited states of NaK with vibrational quantum number $v$ have modified dipole moments $\mu_v$, usually reduced from those of the ground state. These may be estimated using the dipole moment functions of ref.\ \cite{Aymar:2005}. The rotational constant $b_v$ varies with $v$ as well \cite{Russier-Antoine:2000}. The key quantity is $\tilde{b}_v$, which is proportional to $b_v \mu_v^4$. Figure \ref{fig:NaK-scaled:results} shows a calculation for $v=25$, with $b_v \approx 0.85b_0$ and $\mu_v$ estimated as $0.95\mu_0$, corresponding to $\tilde{b}_v=5.7 \times 10^8$. The ratio $b_v/\mu_v$ is smaller for $v=25$ than for $v=0$, so effective shielding occurs at somewhat lower fields. In this case the zeroes in $\alpha$ occur at 6.31 and 6.70 kV/cm, where shielding is very effective ($\gamma\sim 3200$ and $650$, respectively).

\begin{figure}[tbp]
		\includegraphics[width=0.45\textwidth]{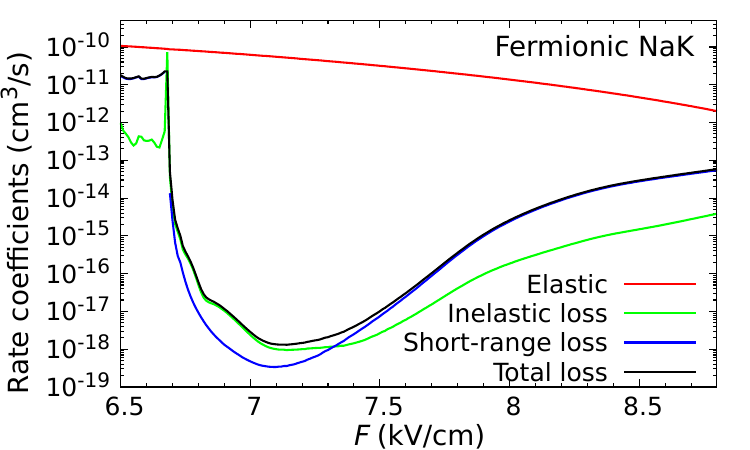}
    \caption{Rate coefficients for two-body elastic collisions and loss processes for ground state fermionic NaK as a function of electric field calculated at collision energy $E_\textrm{coll}/k_\textrm{B}=10$\ nK. }%
    \label{fig:NaK-fermionic:results}
\end{figure}

Fermionic molecules are also of great interest. Schindewolf \emph{et al.}\ ~\cite{Schindewolf:NaK-degen:2022} have recently succeeded in cooling fermionic $^{23}$Na$^{40}$K to quantum degeneracy by evaporative cooling assisted by microwave shielding. Figure \ref{fig:NaK-fermionic:results} shows the calculated rate coefficients for elastic scattering and loss for this system in the presence of a static electric field. The calculation is the same as for bosons, except that symmetrization with respect to exchange is handled differently and the sum over $L$ is taken over odd values. As expected, shielding is even more effective than for bosonic NaK, because s-wave scattering is forbidden and all channels have a centrifugal barrier that itself suppresses close collisions. In this case $\gamma$ reaches $10^7$ at electric fields near 7.1 kV/cm.

\subsection{NaRb and NaCs}
\label{sec:NaRb}

\begin{figure*}[tbp]
	\subfloat[]{
		\includegraphics[width=0.45\textwidth]{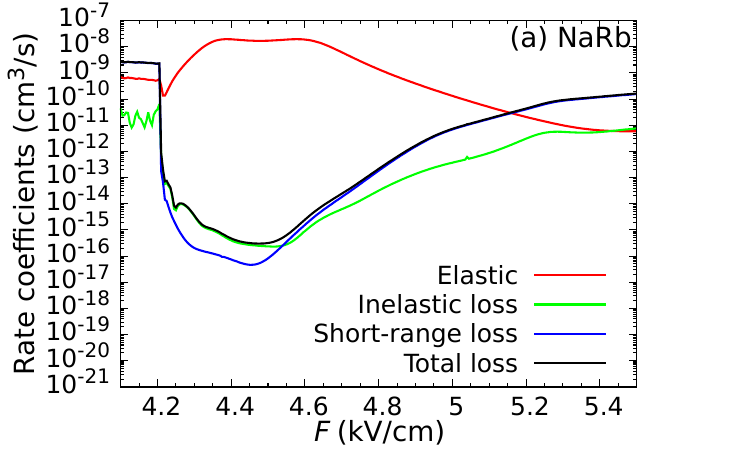}
	}
	\subfloat[]{
		\includegraphics[width=0.45\textwidth]{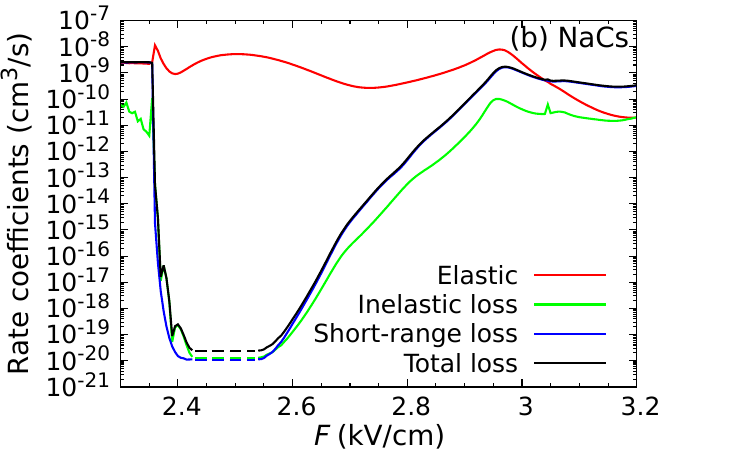}
	}
    %%%%%%%%%%%%%%%%%%%%%%%%%%%%%%%%%%%%second row
    \vspace{-0.8 cm}
	\subfloat[]{
		\includegraphics[width=0.45\textwidth]{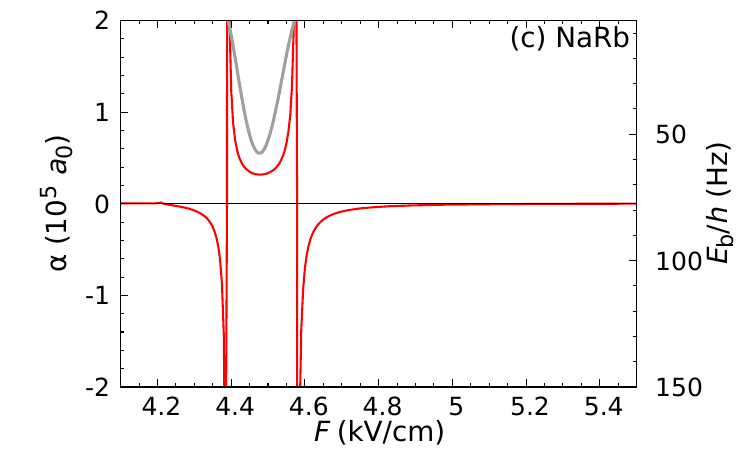}
	}
	\subfloat[]{
		\includegraphics[width=0.45\textwidth]{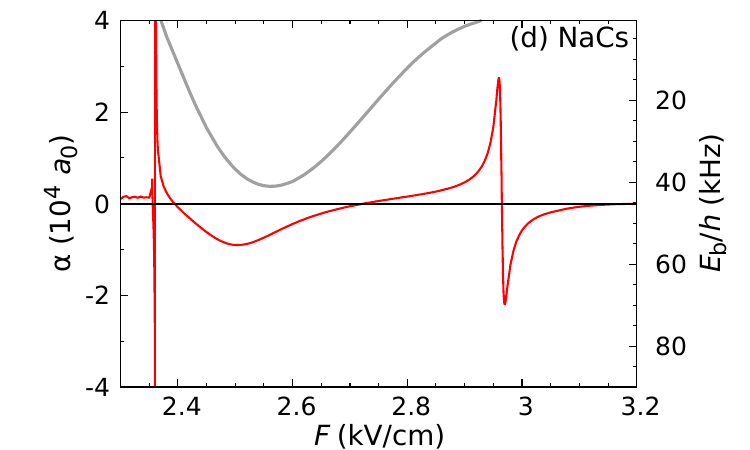}
	}
    \vspace{-0.5 cm}
    \caption{Rate coefficients for two-body elastic collisions and loss processes as a function of electric field for (a) NaRb and (b) NaCs calculated at $E_\textrm{coll}/k_\textrm{B}=10$ nK. For NaCs at 2.43 kV/cm $< F < 2.53$ kV/cm the loss rates are limited by numerical precisions and the dashed lines show upper bounds. Real part $\alpha(k_0)$ of scattering length for (c) NaRb and (d) NaCs. The binding energies $E_\textrm{b}$ of the tetramer bound states are shown as grey lines.}%
\label{fig:NaRb_NaCs:results}
\end{figure*}

Very recently, evaporative cooling has been achieved for bosonic NaRb~\cite{Lin:NaRb:2023} and NaCs~\cite{Bigagli:NaCs:2023} by means of microwave shielding. Here we consider the effect of static-field shielding for these two molecules. NaRb and NaCs are more polar than RbCs and NaK and thus have higher values of $\tilde{b}$, which in turn produces more effective shielding. Figs.\ \ref{fig:NaRb_NaCs:results}a and \ref{fig:NaRb_NaCs:results}c show the calculated rate coefficients for NaRb and NaCs as a function of electric field. The results shown are for $^{23}$Na$^{87}$Rb and $^{23}$Na$^{133}$Cs, but other isotopologues would give results almost identical on the scale of the Figure.
For NaRb at an electric field of 4.5 kV/cm, the rate coefficient for total loss is about 7 orders of magnitude lower than that for elastic collisions. For NaCs at an electric field of 2.5 kV/cm, the corresponding ratio $\gamma$ is at least $10^{10}$ (and is probably substantially more, because our calculated loss rates are limited by numerical precision between 2.43 kV/cm and 2.53 kV/cm).

Figures \ref{fig:NaRb_NaCs:results}b and \ref{fig:NaRb_NaCs:results}d show the real parts $\alpha$ of the low-energy scattering lengths for NaRb and NaCs as a function of electric field. For both these molecules, the long-range attraction is strong enough that the potential well supports a bound state at the center of the shielding region. The well depth decreases towards the edge of the shielding regions, and there are poles in the scattering lengths at the fields where the state crosses threshold. For NaRb, these appear at 4.38 and 4.57 kV/cm, with $\gamma\sim 10^7$ in both cases. For NaCs they appear at 2.37 and 2.97 kV/cm, with $\gamma\sim 10^8$ and 4, respectively. In addition, for NaCs there are zeroes in the scattering length at 2.395 and 2.725 kV/cm, where $\gamma\sim 10^9$ and $10^4$, respectively. These will offer similar opportunities to those afforded by the zeroes in NaK, but in regions of higher shielding and in the presence of a single weakly bound state. There is also good shielding ($\gamma \sim 5000$) at the zero at 2.356 kV/cm, where there is no weakly bound state and 3-body recombination cannot occur. However, this is very close to the field where the pair levels (1,0)+(1,0) and (0,0)+(2,0) cross; as a result the derivative $d\alpha/dF$ is very large, about 4000 $a_0$/(V/cm), so precise field control would be needed to study this region experimentally.

Tuning the electric field across one of the poles will convert pairs of diatomic molecules into weakly bound tetraatomic molecules, termed field-linked states by Avdeenkov and Bohn \cite{Avdeenkov:2003}. This process has been discussed theoretically by Qu\'em\'ener \emph{et al.}\ \cite{Quemener:electroassoc:2023} for the case of microwave shielding, and has been achieved experimentally for fermionic NaK by Chen \emph{et al.}\ \cite{Chen:field-linked-states:2023} by varying the ellipticity of the microwave radiation across an analogous resonance. Qu\'em\'ener \emph{et al.}\ coined the term electroassociation, which seems even more appropriate for association with a static electric field than with a microwave field.

Figures \ref{fig:NaRb_NaCs:results}c and \ref{fig:NaRb_NaCs:results}d show the binding energies of the bound states for NaRb and NaCs as a function of electric field. These are obtained by coupled-channel bound-state calculations \cite{Hutson:CPC:1994}, with dissociation to the open channels prevented by hard-wall boundary conditions at 50 and 20000 $a_0$ \cite{Rowlands:2007}. These states are interesting because they offer a possible route into ultracold cluster chemistry with 4-atom (and perhaps larger) metal clusters. However, the existence of such weakly bound states may enhance 3-body recombination in regions of field where they exist, and thus inhibit evaporative cooling at those fields.

The states of the tetraatomic collision complexes that causes the resonances are formally quasibound rather than bound. They have finite lifetimes, even below threshold, because they can decay through inelastic and short-range loss processes. Under these circumstances there is no actual pole in $\alpha$; instead it undergoes an oscillation of amplitude $\pm a_\textrm{res}/2$ \cite{Hutson:res:2007}. Here $a_\textrm{res}$ is very large, about $4\times10^{10}$ and $3\times10^{10}$ $a_0$ for the low-field and high-field features in NaRb and $4\times10^{7}$ and 50000 $a_0$ for the corresponding features in NaCs. This still provides a very wide range of tuning of $\alpha$, and for simplicity we refer to the features as poles in the current work. The lifetime $\tau$ of the state below threshold is
\begin{equation}
\tau = \frac{\hbar}{\Gamma_\textrm{loss}},
\end{equation}
where $\Gamma_\textrm{loss}$ is the energy width of the quasibound state. We have obtained $\Gamma_\textrm{loss}$ by locating the resonant signature in the eigenphase sum from a coupled-channel scattering calculation below the incoming threshold, as described in ref.\ \cite{Frye:quasibound:2020}.
At the field where the binding energy is greatest, we obtain lifetimes greater than 100~ms for NaRb and 1000~s for NaCs.

\subsection{KAg and CsAg}
\label{sec:KAg}

\begin{figure*}[tbp]
	\subfloat[]{
		\includegraphics[width=0.45\textwidth]{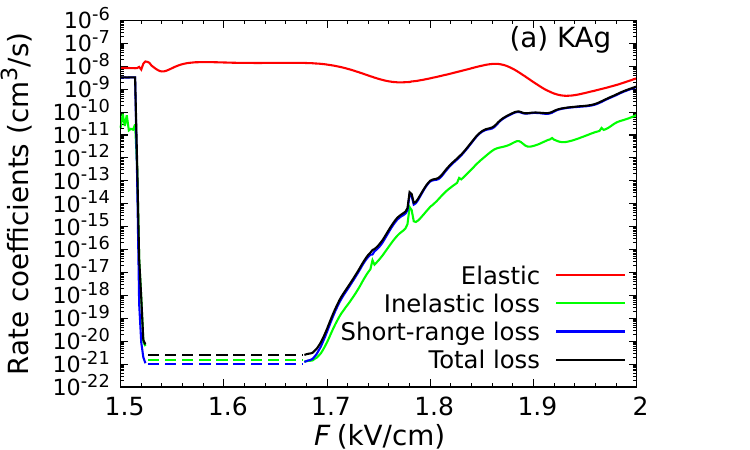}
	}
	\subfloat[]{
		\includegraphics[width=0.45\textwidth]{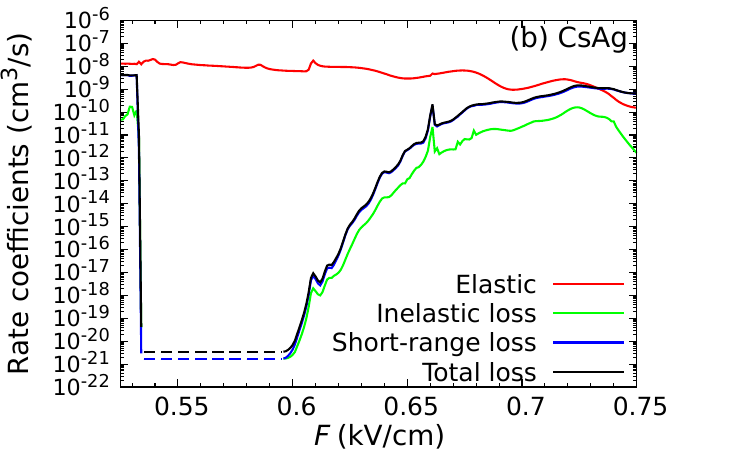}
	}
    %%%%%%%%%%%%%%%%%%%%%%%%%%%%%%%%%%%%second row
    \vspace{-0.8 cm}
	\subfloat[]{
		\includegraphics[width=0.45\textwidth]{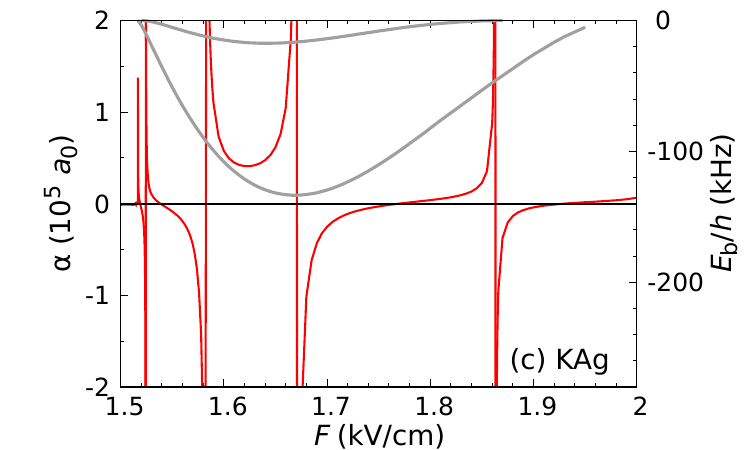}
	}
	\subfloat[]{
		\includegraphics[width=0.45\textwidth]{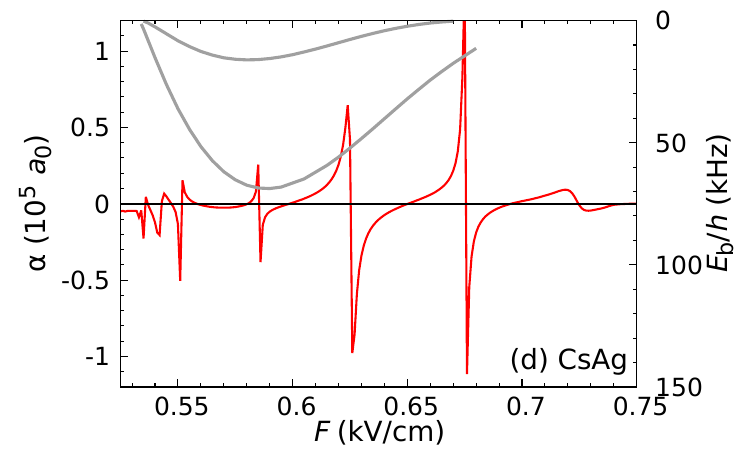}
	}
    \vspace{-0.5 cm}
    \caption{Rate coefficients for two-body elastic collisions and loss processes as a function of electric field for (a) KAg and (b) CsAg calculated at $E_\textrm{coll}/k_\textrm{B}=10$ nK. Loss rates at electric fields within the range of 1.52 to 1.68 kV/cm for KAg and 0.547 to 0.559 kV/cm for CsAg are limited by numerical precision and the dashed lines show upper bounds. Real part $\alpha(k_0)$ of the scattering length for (c) KAg and (d) CsAg. These systems support up to 3 and 4 tetramer bound states, respectively; the binding energies of the deeper states are shown as grey lines in panels (c) and (d).}%
\label{fig:KAg_CsAg:results}
\end{figure*}

Diatomic molecules formed from an alkali-metal atom and Cu or Ag are isoelectronic with alkali-metal dimers, but are expected to have larger dipole moments~\cite{Smialkowski:2021} and correspondingly larger values of $\tilde{b}$. In this subsection we investigate electric-field shielding of KAg and CsAg. We consider bosonic molecules and use weighted-average atomic masses for K and Ag, based on their isotopic abundances. Figures \ref{fig:KAg_CsAg:results}a and \ref{fig:KAg_CsAg:results}b show the resulting rate coefficients as a function of electric field. The rate coefficient for total loss is at least 12 orders of magnitude lower than that for elastic scattering.

Figures \ref{fig:KAg_CsAg:results}c and \ref{fig:KAg_CsAg:results}d show the real parts $\alpha$ of the scattering lengths for KAg and CsAg. There are 3 bound states for KAg in the middle of the shielding region and 4 for CsAg. These bound states enter and leave the potential well as a function of electric field. However, the outermost poles in each case are strongly suppressed by loss processes, and the corresponding states will be short-lived in this region. Nevertheless, there are multiple poles and zeroes for each system in the range of fields where shielding is effective, so the scattering length is highly tunable. For KAg, for example, the scattering length passes through zero at $F=1.54$ kV/cm, where $\gamma > 10^{12}$.

\section{Conclusions}
\label{sec:conclusions}

We have investigated the use of a static electric field to shield polar molecules from short-range collisions that cause trap loss. The shielding is due to dipole-dipole interactions between the molecules, which can be made repulsive by engineering a near-degeneracy between two pair states with different rotational quantum numbers \cite{Avdeenkov:2006, Wang:dipolar:2015}.  We have explored the behavior of a representative range of ultracold polar molecules of current experimental interest: RbCs, NaK, NaRb, NaCs, KAg and CsAg. Static-field shielding has distinct differences from microwave shielding \cite{Karman:shielding:2018, Karman:shielding-imp:2019, Lassabliere:2018}, and may be advantageous in some cases. For example, microwave shielding will be difficult for molecules with low rotational constants, such as RbCs and CsAg, which require impractically large electrodes to generate near-resonant microwaves.

Important differences between molecules arise because the dipole-dipole interaction produces a long-range attraction in the effective potential (adiabat) for s-wave scattering. This attraction exists at intermolecular distance \emph{outside} the repulsion that produces shielding. Without the attraction, the repulsive core of the shielding interaction would produce a large positive scattering length. However, the attraction adds a negative contribution and allows substantial tuning of the scattering length with electric field. If the attraction is strong enough, the resulting long-range well can support weakly bound states.

The attractive and repulsive parts of the interaction are affected differently by applied field. In many cases, the same scattering length can be achieved with several different applied fields, corresponding to different space-fixed dipole moments. This will make it possible to investigate many-body properties as independent functions of scattering length and dipole moment.

For RbCs, microwave shielding is impractical, at least on the lowest rotational transition, because the required wavelength is very long and the antennas needed to generate the radiation are very large. In addition, very strong microwave fields are needed. However, static-field shielding occurs at easily accessible electric fields (near 2.7 kV/cm) and is quite effective, with a ratio of 2-body elastic to loss rates $\gamma$ of about $10^3$ at 10~nK and $10^4$ at 1~$\mu$K. For RbCs, the long-range attraction is sufficient to reduce the scattering length to moderate positive values (${\sim}200\ a_0$) that are very favorable for evaporative cooling towards quantum degeneracy. Such values are also favorable for forming a dipolar Bose-Einstein condensate (BEC) that will not collapse \cite{Ronen:2006}. Moreover, the absence of a long-range bound state means that 3-body recombination cannot occur.

For NaK, static-field shielding requires larger fields than for RbCs (about 7 kV/cm) and is even more effective than for RbCs ($\gamma > 10^4$ at 10~nK). The long-range attractive well is deeper than for RbCs, but not deep enough to support a bound state. The scattering length is negative in the region where shielding is strongest, but there are zeroes in the scattering length as a function of field, including one at 6.90 kV/cm where $\gamma$ is ${\sim}2500$. This may provide opportunities to study the stability of a dipolar BEC as the scattering length is tuned from positive to negative values. In excited vibrational states of NaK, with smaller rotational constants, the zeroes in the scattering length move towards the fields where shielding is most effective. Once again, the absence of a long-range bound state precludes 3-body recombination.

For NaRb and NaCs, the attractive well is deep enough to support a single bound state at the center of the shielding region. This state is very shallow for NaRb (binding energy $E_\textrm{b}$ up to 60~Hz$\times h$), but rather deeper for NaCs ($E_\textrm{b}$ up to 40 kHz$\times h$). For each molecule there are two poles in the scattering length as the bound state crosses the incoming threshold, and for NaCs there are additional zeroes in the scattering length between the poles, in the region where a long-range bound state exists. The poles may allow electroassociation to form tetraatomic molecules and the study of Efimov physics in dipolar systems, while the additional zeroes provide opportunities to study the stability of dipolar BECs in the presence of 3-body recombination.

For KAg and CsAg, and other molecules with very large dipoles, the attractive well is deep enough to support multiple bound states. There are multiple poles in the scattering length where the states cross the incoming threshold and multiple zeroes between the poles. Such systems will provide opportunities to study the behavior of dipolar BECs under a wide range of conditions, and to study the effects of scattering length and space-fixed dipole independently.

In conclusion, static-field shielding provides an important tool for shielding dipolar molecules from destructive collisions and for controlling their scattering length. It is complementary to microwave shielding, with some similarities but also important differences. Different molecules show quite different behaviors, and static-field shielding offers access to much new physics that is so far unexplored.\\

\section*{Rights retention statement}

For the purpose of open access, the authors have applied a Creative Commons Attribution (CC BY) licence to any Author Accepted Manuscript version arising from this submission.

\section*{Data availability statement}

Data supporting this study are openly available from Durham University \cite{DOI_alkali_dc_shielding}.

\section*{Acknowledgement}
We are grateful to Matthew Frye and Ruth Le Sueur for valuable discussions.
This work was supported by the U.K. Engineering and Physical Sciences Research Council (EPSRC) Grant Nos.\ EP/W00299X/1, and EP/V011677/1.

\bibliographystyle{../long_bib}
\bibliography{../all,Alk_dc_shieldingData}
\end{document}